\documentclass[aps,prl,twocolumn,showpacs,superscriptaddress]{revtex4}
\usepackage{graphicx}
\begin{document}



\title{Muon Spin Rotation Measurement of the Magnetic Field Penetration Depth in Ba(Fe$_{0.93}$Co$_{0.07})_2$As$_2$: Evidence for Multiple Superconducting Gaps}
     \author{T.~J.~Williams}
     \affiliation{Department of Physics and Astronomy, McMaster University, Hamilton, Ontario, Canada, L8S 4M1}
     \author{A.~A.~Aczel}
     \affiliation{Department of Physics and Astronomy, McMaster University, Hamilton, Ontario, Canada, L8S 4M1}
     \author{E.Baggio-Saitovitch}
 \affiliation{Centro Brasilieriro de Pesquisas Fisicas, Rua Xavier Sigaud 150 Urca, CEP 22290-180 Rio de Janeiro, Brazil}
     \author{S.~L.~Bud'ko}
     \affiliation{Department of Physics and Astronomy and Ames Laboratory, Iowa State University, Ames, Iowa 50011, USA}
     \author{P.~C.~Canfield}
     \affiliation{Department of Physics and Astronomy and Ames Laboratory, Iowa State University, Ames, Iowa 50011, USA}
     \author{J.~P.~Carlo}
     \affiliation{Department of Physics, Columbia University, New York, NY 10027, USA}
     \author{T. Goko}
     \affiliation{Department of Physics and Astronomy, McMaster University, Hamilton, Ontario, Canada, L8S 4M1}
     \affiliation{Department of Physics, Columbia University, New York, NY 10027, USA}
     \affiliation{TRIUMF, Vancouver, British Columbia, Canada, V6T 2A3}
     \author{J.~Munevar}
     \affiliation{Centro Brasilieriro de Pesquisas Fisicas, Rua Xavier Sigaud 150 Urca, CEP 22290-180 Rio de Janeiro, Brazil}
         \author{N.~Ni}
     \affiliation{Department of Physics and Astronomy and Ames Laboratory, Iowa State University, Ames, Iowa 50011, USA}
     \author{Y.~J.~Uemura}
     \affiliation{Department of Physics, Columbia University, New York, NY 10027, USA}
     \author{W.~Yu}
     \affiliation{Department of Physics and Astronomy, McMaster University, Hamilton, Ontario, Canada, L8S 4M1}
     \author{G.~M.~Luke}
     \altaffiliation{author to whom correspondences should be addressed: E-mail:[luke@mcmaster.ca]}
     \affiliation{Department of Physics and Astronomy, McMaster University, Hamilton, Ontario, Canada, L8S 4M1}
     \affiliation{Canadian Institute of Advanced Research, Toronto, Ontario, Canada, M5G 1Z8}

\date{\today}

\begin{abstract}
We have performed transverse field muon spin rotation measurements of single crystals of Ba(Fe$_{0.93}$Co$_{0.07})_2$As$_2$ with the
applied magnetic field along the $\hat{c}$ direction. Fourier transforms of the measured spectra reveal an anisotropic lineshape characteristic of an Abrikosov vortex lattice.
We have fit the $\mu$SR spectra to a microscopic model in terms of the penetration depth $\lambda$ and the Ginzburg-Landau parameter $\kappa$.
 We find that as a function of temperature, the penetration depth varies
more rapidly than in standard weak coupled BCS theory. For this reason we first
 fit the temperature dependence to a power law where the power varies from 1.6 to 2.2 as the field changes from 200G to 1000G.
Due to the surprisingly strong field dependence of the power and the superfluid density we proceeded to
 fit the temperature dependence to a two gap model, where the size of the two gaps is field independent.  From this model,
 we obtained  gaps of $2\Delta_1=3.7k_BT_c$ and $2\Delta_2=1.6k_BT_c$, corresponding
to roughly 6~meV and 3~meV respectively.
\end{abstract}

\maketitle

Recently, a new family of high temperature superconductors were discovered\cite{Kamihara_08} based on layers of FeAs. Several groups of these structures have been found, including the so-called
1111 compounds, such as LaFeAsO, and the 122 compounds, such as BaFe$_2$As$_2$. Superconductivity was discovered in these systems
 by hole-doping, such as partially replacing oxygen by fluorine in LaFeAsO or partially substituting potassium for barium in BaFe$_2$As$_2$.
  The resulting superconductors have remarkably high T$_c$'s, for example, up to 56K for LaFeAs(O$_{1-x}$F$_x$)\cite{Sefat_08a,Chen_08,Liu_08} and up to 38K for (Ba$_{1-x}$K$_x$)Fe$_2$As$_2$\cite{Rotter_08,Sasmal_08,Ni_08a}.

At room temperature the parent compounds have a tetragonal structure, which then undergoes an orthorhombic distortion somewhat below 200~K. Either at or just below the structural phase
transition, there is a magnetic phase transition to an antiferromagnetic (AF) ground state. With doping, the AF transition is supressed, and superconductivity emerges.
The proximity to AF magnetic order has led to a belief that Fe spin fluctuations are important for developing the superconducting ground state\cite{Haule_08}.
 More recently, it was found that superconductivity can be induced in these systems through electron-doping through substitution of iron in the FeAs layers; both La(Fe$_{1-x}$Co$_x$)AsO\cite{Sefat_08b} and Ba(Fe$_{1-x}$Co$_{x}$)$_2$As$_2$\cite{Sefat_08c,Ni_08} become superconducting with Tc's as high as 14 and 23~K respectively. In contrast to the cuprates, the iron pnictides are apparently surprisingly robust against in-plane disorder.
 
At this point there is no consensus on the pairing symmetry in pnictides. In the 1111 system most experiments indicate a fully gapped Fermi surface, although the presence of magnetic rare earth atoms
and the lack of large single crystals complicates the interpretation\cite{Mazin_09}. In the 122 systems the situation is also unclear. Tunnel diode resonator measurements\cite{Gordon_08a,Gordon_08b}
of Ba(Fe$_{1-x}$Co$_x$)$_2$As$_2$
exhibit power law temperature dependences for the penetration depth $\Delta\lambda(T)\sim T^n$ with $n$ between 2 and 2.5 which have been interpreted in terms of gap nodes.
Similar measurements of Ba$_{1-x}$K$_x$Fe$_2$As$_2$\cite{martin_09} exhibited a power law $n$ of about 2 for low (T/T$_c\leq0.25$ and were fit with a two gap s-wave
 model with somewhat unphysical parameters. Muon spin rotation
measurements of single crystal Ba$_{1-x}$K$_x$Fe$_2$As$_2$ exhibited phase separation into magnetic and superconducting regions\cite{Aczel_08,Goko_09}, with different temperature dependences for
the superfluid density ($n_s\propto 1/\lambda^2$) depending on the value of $x$. NMR 1/T$_1$ measurements\cite{Ning_08} show no Hebel-Slichter peak below T$_c$ which generally indicates
non-s-wave pairing, whereas the $^{59}$Co and $^{75}$As Knight shifts decrease below T$_c$ for fields both along the c-axis and in the a-b plane; behavior which is consistent with s-wave pairing.

Muon spin rotation ($\mu$SR) is a powerful local microscopic tool for characterizing the magnetic properties of materials, in superconducting or other states. A thorough description of the application of $\mu$SR to studies of superconductivity can be found elsewhere\cite{Sonier_07}. In a transverse field (TF) $\mu$SR experiment, spin polarized positive muons are
implanted one at a time into a sample.
 The muon spins precess around the local magnetic field, and decay into a positron, which is preferentially ejected along the direction of the muon spin at the time of decay (as well as two neutrinos which are not detected). In the presence of a vortex lattice, the spatial variation of the magnetic field distribution results in a dephasing of the muon spin polarization and a relaxation of the precession signal. A Fourier transform of the spin polarization function essentially reveals the field distribution which exhibits a characteristic Abrikosov lineshape. The lineshape (or equivalently the relaxation function in the time domain) depends on the lattice geometry, magnetic field penetration depth $\lambda$, coherence length, $\xi$, and the amount of lattice disorder. As a result, careful analysis of the relaxation
 function allows these microscopic parameters to be determined in the vortex state. Such measurements demonstrated the presence of gap nodes characteristic of d-wave pairing
 in high quality single crystals of YBa$_2$Cu$_3$O$_{6.97}$\cite{Sonier_07}. In ceramic samples this anisotropic lineshape is generally not
 observed, rather the broadened line is generally well described by a gaussian distribution; however, the width of this distribution (the Gaussian relaxation rate) $\sigma$ has been shown to be
 proportional to the superfluid density $\sigma\propto n_s\propto 1/\lambda^2$\cite{Brandt_88,Maisuradze_09}. Previous studies of cuprates found that extrinsic effects in ceramics can result in
 the correct temperature dependence of the superfluid density being masked; for this reason, reliable measurements of the superfluid density require the use of single crystals and the observation of
 an anisotropic lineshape characteristic of a vortex lattice.

A single crystal of Ba(Fe$_{0.93}$Co$_{0.07})_2$As$_2$ was grown from self flux as described in detail elsewhere\cite{Ni_08}.
This crystal of roughly 1cm$^2$ area was mounted in a helium gas flow cryostat on the M20 surface muon beamline at TRIUMF, using a low background arrangement such that only positrons originating from muons landing in the specimens were
collected in the experimental spectra.

 Previous $\mu$SR work on (Ba,K)Fe$_2$As$_2$ found regions of phase-seperated magnetic order, with spontaneous muon precession in ZF-$\mu$SR spectra \cite{Goko_09}. In
 a recent ZF-$\mu$SR study of Ba(Fe$_{0.9}$Co$_{0.1})_2$As$_2$\cite{Bernhard_09}, weak, temperature dependent relaxation was also observed. The presence of relaxation due to magnetism largely precludes
 detailed analysis of the properties of the vortex lattice and therefore it is important to check that such relaxation is absent. We performed ZF-$\mu$SR measurements at temperatures both above T$_c$ and at T$=2$~K, finding that the spectra were identical, showing only weak temperature independent relaxation, such as would be expected for nuclear dipole moments.
\begin{figure}[]
\includegraphics[angle=90, width=\columnwidth]{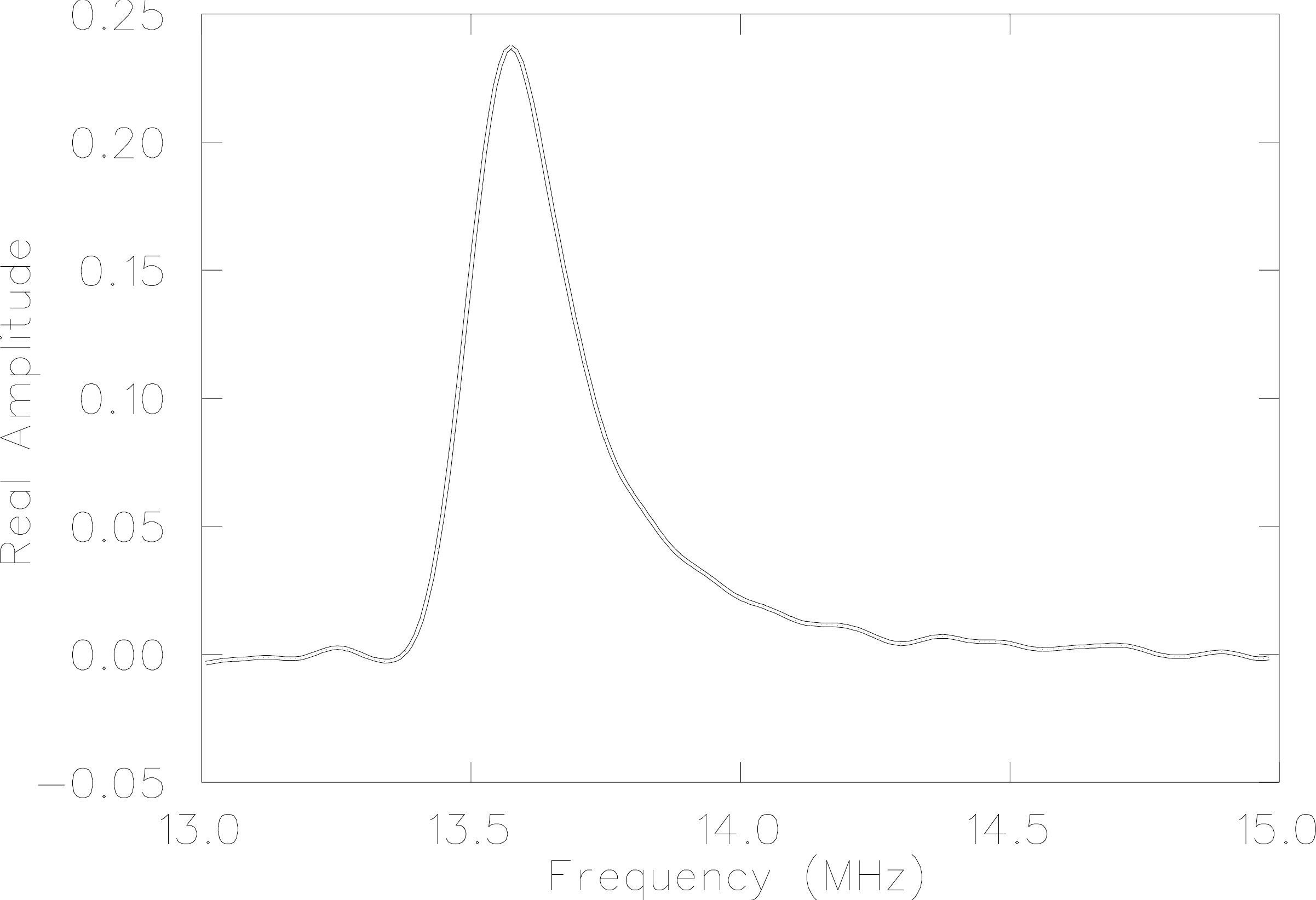}
\caption{\label{fig1}
FFT of the TF $\mu$SR signal in B$_{ext}$=0.1~T at T=1.7~K. The anisotropic lineshape is characteristic of a well-ordered vortex lattice.}
\end{figure}

Fig. 1 shows a fast Fourier transform (FFT) of the TF $\mu$SR signal measured in an applied field of 0.1~T at T$=1.6$~K. The anisotropic lineshape is characteristic of an Abrikosov vortex lattice and indicates the presence of at least
locally well-ordered vortices within the superconducting state. The lower cutoff in the lineshape corresponds to muons landing at the center of the lattice (furthest from vortex cores), the maximum in the lineshape comes from muons at the saddle point midway between vortices while the cutoff at high field comes from muons at the vortex cores. The overall width of the lineshape is dominated by the magnetic penetration depth $\lambda$, while the high field cutoff mainly reflects the coherence length $\xi$. The high field cutoff is most evident in the highest applied fields which is furthest from the London limit where the field would actually diverge at the vortex cores, giving a high field tail with no cutoff.
 
\begin{figure}
\includegraphics[angle=90,width=\columnwidth]{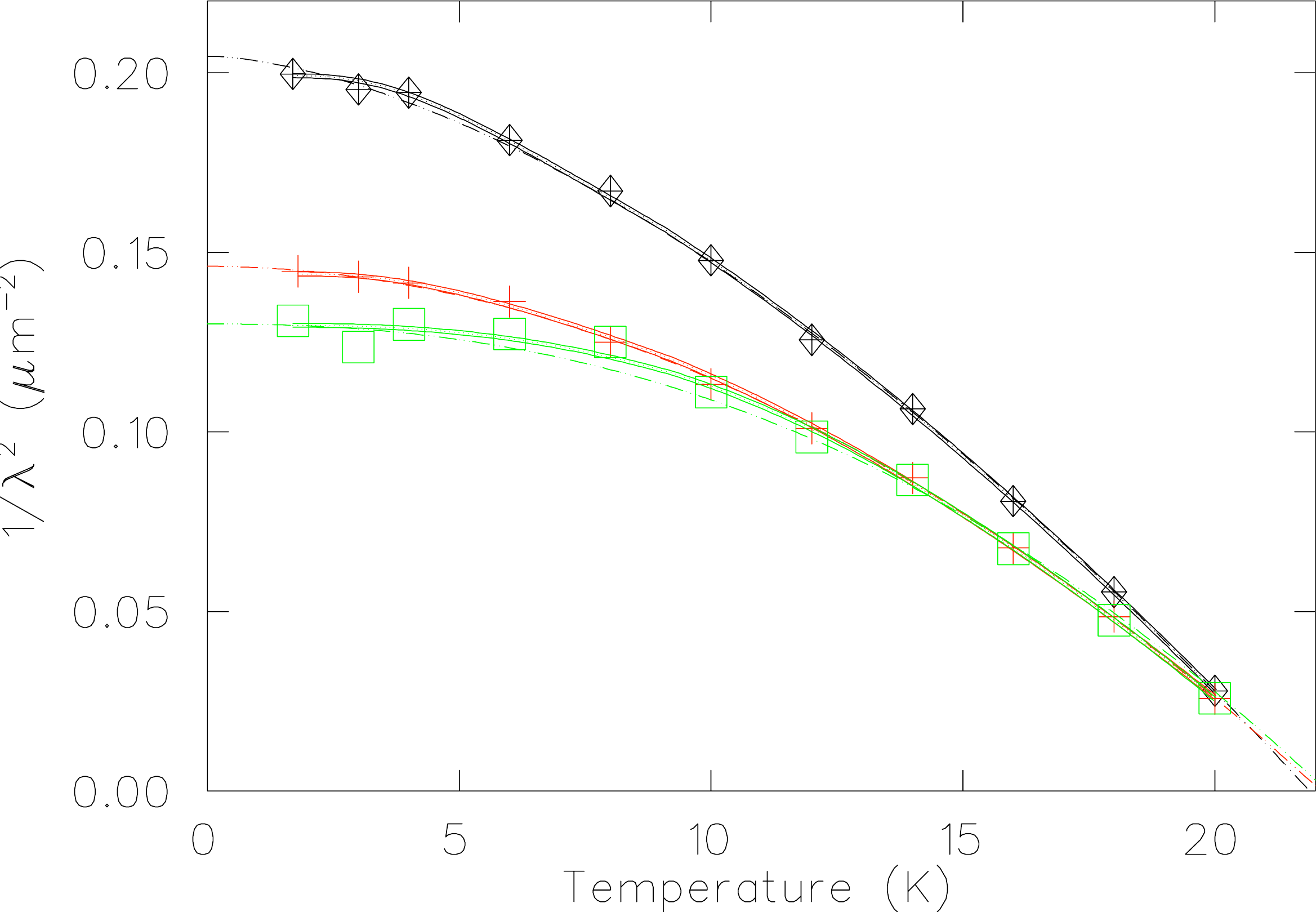}
\caption{\label{super_fluid}
(Color online) Combined two-gap fit (solid line) and power law fit (dashed line) to the superfluid density measured from the TF $\mu$SR measurements at 0.02~T (black diamonds), 0.05~T (red crosses) and 0.1~T (green squares).}
\end{figure}

The data was analyzed by simultaneously fitting several runs to an analytical Ginzburg-Landau model\cite{Sonier_07} for an applied field of 0.1~T, to determine the Ginzburg-Landau parameter, $\kappa=\lambda/\xi$. We found $\kappa=44$; this value was held fixed for the remainder of the analysis, noting that measurements at lower fields are not sensitive to $\kappa$ as long as it remains large.  Small-angle neutron scattering (SANS) measurements~\cite{eskildsen_09}  of vortices in applied fields above 0.2~T found a highly disordered vortex arrangement.  We included the effects of vortex lattice disorder in our analysis via an additional gaussian broadening of our  $\mu$SR spectrum\cite{Brandt_88, Riseman_95}, where we assumed that this broadening was proportional to $1/\lambda^2$ as observed in previous studies of cuprates and other high $\kappa$ superconductors~\cite{Sonier_07}.  Our fitted result shows that the rms deviation of the vortex positions ($\langle s^2\rangle^{1/2}$)
relative to  the vortex separation ($L_0$) was  $\langle s^2\rangle^{1/2}/L_0\approx2.4$~\% at 1.7~K in H$_{ext}=0.1~T$.  This relatively small amount of disorder as seen in the field distribution is somewhat in contrast with the SANS results.  However, the disorder is generally greater in higher fields (as used in the SANS study) and may also reflect the fact that $\mu$SR, as a real space probe is less affected by a loss of true long range order than is a reciprocal space probe such as SANS.  We note that Bitter decoration measurements~\cite{eskildsen_09} (in low fields) provide evidence for at least regions of well ordered vortices.
 At this point, we were able  to fit the rest of the parameters in the $\mu$SR signal, including the penetration depth, $\lambda(H,T)$. Results of this analysis for $1/\lambda^2\propto n_s$ (where $n_s$ is the superfluid density)
are shown in Fig.~\ref{super_fluid} for applied fields of 0.1~T, 0.05~T and 0.02~T where it can be clearly seen that both the temperature dependence and zero temperature value $n_s(T\rightarrow0)$ depend on the applied field.

The temperature dependence of the superfluid density reflects how easily thermal fluctuations are able create quasiparticles. In conventional weak coupling BCS theory one finds that the low temperature behaviour of $n_s$ becomes exponentially flat, whereas the presence of gap nodes is reflected in a low temperature power law behaviour (such as T-linear in the case of the d-wave cuprates). We first fit
to a power law, $n_s(T)=n_s(0)[1-(T/T_c)^{p}]$. The results of this fit are indicated by the dashed lines in Fig. ~\ref{super_fluid}. We found that the fitted power increases with increasing field, from 1.62$\pm$0.03 at 0.02~T to 2.27$\pm$0.03 at 0.1~T while the superfluid density at $T=0$ decreased with field, from 0.2047$\pm$0.0015 $\mu m^{-3}$ to 0.1301$\pm$0.0005 $\mu m^{-3}$. We note that this model provides a fairly good fit to the data, but the strong field dependence of both the superfluid density and the power are somewhat surprising for a range of fields much less than H$_{c2}$\cite{Ni_08} which encourages us to
try different possible models to characterize the superfluid density.

ARPES measurements of the iron pnictides have demonstrated that the Fermi surface has multiple sheets and that there are at least two different superconducting gaps on these bands
and as a result, a multi-band model of the superconducting state has been proposed for these systems\cite{Kondo_08,Liu_08,Hunte_08,Mu_08,Jaroszynski_08,Terashima_08}.
 Multigap superconductivity has been seen in a number of systems including
MgB$_2$ \cite{Nagamutsu_01}, which is an s-wave superconductor with a T$_C$ of 39K and where theoretical calculations predict that the smaller gap is an induced gap associated with the 3D bands, and the larger gap is associated with the superconducting 2D bands\cite{Liu_08a}. The possibility of other materials exhibiting multi-band superconductivity was realized with the discovery of the two-band superconductor NbSe$_2$ by ARPES measurements\cite{Yokoya_01}. This situation has been clearly observed in $\mu$SR measurements\cite{Callaghan_05,Sonier_97}. Like MgB$_2$, this material had two gaps of very different sizes, which reside on two sets of Fermi surface sheets\cite{Yokoya_01}. Multi-band superconductivity has been proposed in many other materials, including CeCoIn$_5$\cite{Barzykin_07,Tanatar_05}, BaNi$_2$P$_2$\cite{Terashima_09} and PrOs$_4$Sb$_{12}$\cite{MacLaughlin_07}.

We next fit the superfluid density to a phenomenological two-gap model\cite{Bouquet_01b,Ohishi_03} which has been employed in a previous $\mu$SR study of LaFeAs(O,F), Ca(Fe,Co)AsO, and (Ba,K)Fe$_2$As$_2$.\cite{Takeshita_08}

\begin{equation}n_s(T)=
n_s(0)-w\cdot\delta n_s(\Delta_1,T)-(1-w)\cdot\delta n_s(\Delta_2,T)
\label{eq1}
\end{equation}

where $w$ is the relative weight for the first gap, $\Delta_1$. Here, the gap functions are given by,

\begin{equation}\delta n(\Delta,T)=
\frac{2n_s(0)} {k_BT}\int_{0}^{\infty}f(\epsilon,T)\cdot[1-f(\epsilon,T)]d\epsilon
\label{eq2}
\end{equation}

where $f(\epsilon,T)$ is the Fermi distribution given by,

\begin{equation}f(\epsilon,T)=
(1+e^{\sqrt{\epsilon^2+\Delta(T)^2}/k_BT})^{-1}
\label{eq3}
\end{equation}

Here, $\Delta_i$ ($i=1$ and $2$) are the energy gaps at $T=0$, and $\Delta_i(T)$ were taken to follow the standard BCS temperature dependence.
The size of the gaps, $\Delta_1$ and $\Delta_2$, and $T_c$ were fit globally, while $n_s(0)$ and the weighting factor, $w$, were allowed to be field-dependent.

The results of this analysis are shown by the solid lines in Fig. ~\ref{super_fluid}. It can be seen that this gives a very good fit, with a $\chi^2$ that are approximately half of that for the power law fit. Based on other experimental support for a two-gap model, this gives support to our choice of fit.

From the fit, we obtained the values of the gaps $2\Delta_1/k_BT_c=3.768$ and $2\Delta_2/k_BT_c=1.565$, and $T_c=22.1\pm 0.2K$. The larger of the two gap values is quite close to the BCS value, while the smaller gap is roughly half the BCS gap. These gaps are lower than has been reported for other iron pnictide compounds, which presumably indicates that the strength of the superconducting pairing varies from system to system.

The relative weighting factor for the larger gap increases with $B_{app}$, from $w=0.655(7)$ at 0.02~T, $w=0.766(6)$ at 0.05~T to $w=0.909(4)$ at 0.1~T. This indicates that the applied field acts to weaken superconductivity on the bands with the smaller gap. The superfluid density at $T=0$, $n_s(0)$, decreases with $B_{app}$, giving $n_s(0)=0.199(1), 0.1441(4), 0.130(1)$ for these fields respectively.

\begin{figure}
\includegraphics[width=\columnwidth,angle=0]{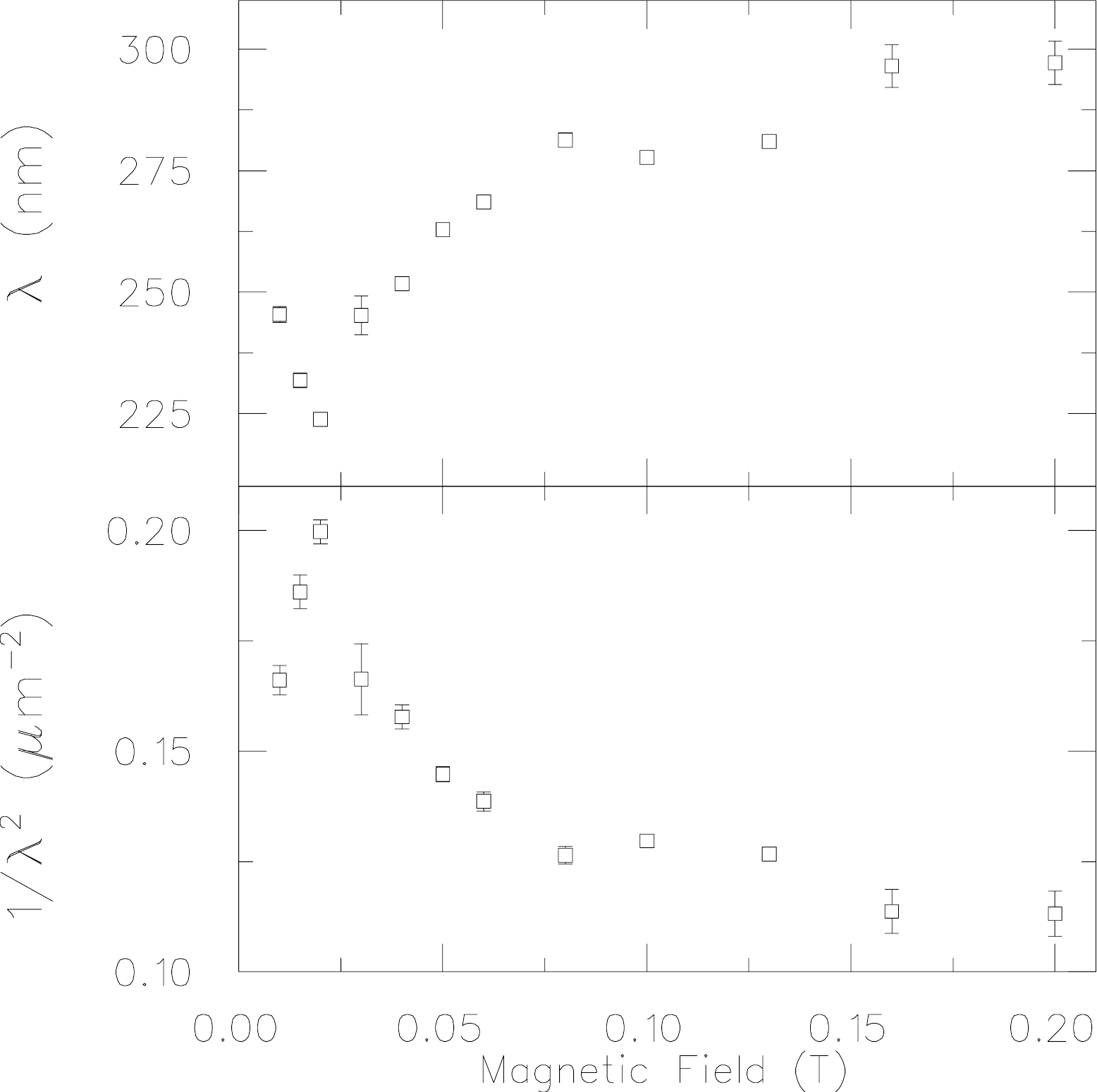}
\caption{\label{field_dep}
(top) Magnetic field dependence of the magnetic field penetration depth in Ba(Fe$_{0.93}$Co$_{0.07})_2$As$_2$ measured at T$=1.7$~K. (Bottom)$1/\lambda^2 \propto$ superfluid density, $n_s$, measured at T=1.7K. }
\end{figure}

In addition to measuring the full temperature dependence of the penetration depth for 3 applied fields, we also studied the magnetic field dependence of $n_s$ at T=1.7K (following field-cooling through T$_C$ for each applied field). We simultaneously fit pairs of $\mu$SR spectra at T$=1.7$~K and in the normal state using the same procedure as before to obtain the penetration depth at T$=1.7$~K. The results of this analysis are shown in Fig.~\ref{field_dep}. We can see that the superfluid density shows a small peak near 0.02~T, then decreases with increasing field,  which is in agreement with the trend found in our full temperature scans. The peak at low fields may reflect the proximity of the applied field to the lower critical field, which is about 0.007~T at Y$=5$~K\cite{Gordon_08a}.
The penetration depth tends to a constant value of around 300~nm for  higher fields. Recalling the field dependence of the relative gap weighting factor $w$, this high field saturation behaviour presumably reflects the loss of superconductivity on the bands with the smaller gap.

In conclusion, we have performed transverse field $\mu$SR measurements in the mixed state of single crystal Ba(Fe$_{0.93}$Co$_{0.07})_2$As$_2$.  Employing an analytic Ginzburg-Landau theory to obtain the penetration depth $\lambda$ and Ginzburg-Landau parameter $\kappa$ we find that the superfluid density $n_s\propto1/\lambda^2$ can be well-described by a s-wave two gap model where 
the corresponding field-independent gaps are $2\Delta_1=3.7k_BT_c$ and $2\Delta_2=1.6k_BT_c$ respectively.

\begin{acknowledgments}
We thank Jeff Sonier for sharing his computer program for evaluating vortex lattice disorder.  P.C.C. and G.M.L. thank R. Prozorov for useful discussions.
We appreciate the hospitality of the TRIUMF Centre for Molecular and Materials Science where the majority of these experiments were performed. A.A.A. is supported by an NSERC
CGS fellowship. Research at McMaster University is supported by NSERC and CIFAR.  Work at Columbia was supported by NSF-DMR-0502706 and NSF-DMR-0806846.
Work at Ames Laboratory was supported by the Department of Energy, Basic Energy Sciences under Contract No. 
DE-AC02-07CH11358. \\
\end{acknowledgments}

\vfill \eject
\end{document}